\documentclass[aps,prl,twocolumn,groupedaddress]{revtex4}

\usepackage[latin1]{inputenc}
\usepackage{amsmath}
\usepackage{verbatim}
\usepackage{amssymb}
\usepackage{bbm}
\usepackage{graphicx}

\newcommand{\approptoinn}[2]{\mathrel{\vcenter{
  \offinterlineskip\halign{\hfil$##$\cr
    #1\propto\cr\noalign{\kern2pt}#1\sim\cr\noalign{\kern-2pt}}}}}

\begin{document}


\title{Auditory power-law activation-avalanches exhibit a fundamental computational ground-state}

\author{Ruedi Stoop and Florian Gomez}
\affiliation{Institute of Neuroinformatics and Institute of Computational Science, University of Zurich and ETH Zurich, Winterthurerstr. 190, 8057 Zurich, Switzerland}


\date{\today}

\begin{abstract}
The cochlea provides a biological information-processing paradigm that we only begin to understand in its full complexity. Our work reveals an interacting network of strongly nonlinear dynamical nodes, on which even simple sound input triggers subnetworks of activated elements that follow power-law size statistics ('avalanches'). From dynamical systems theory, power-law size distributions relate to a fundamental ground-state of biological information processing. Learning destroys these power laws. These results strongly modify the models of mammalian sound processing and provide a novel methodological perspective for understanding how the brain processes information. 
\end{abstract}
\pacs{}
\maketitle

The mammalian neocortex has recently been suggested to host neuronal activation {\em avalanches} \cite{Plenz,Plenz2,Beggs3} of power-law distributed size. Evidence from simulated synchronizing phase oscillator networks \cite{Boccaletti,Boccaletti-masterstabilityfunction} and other models \cite{Eurich,Herrmann,Hans,Niebur} supports this suggestion; however, the methods underlying the experimental results are delicate, and most of the models use strongly simplified neuronal dynamics or implement particular balance conditions. Because avalanches were related to a critical state in cortex, which has opened a discussion on the significance of such a potential state regarding information propagation and processing capacity \cite{Haldeman, Beggs3,Dante1,Eurich,Arcangelis,Hans}, additional supporting evidence would be desirable. 

The mammal's hearing sensor, the {\em cochlea}, can be seen as an information-processing antecedent of the neocortex that avoids most of the mentioned experimental and theory difficulties. In essence, the cochlea consists of a continuous, fluid-embedded stiffness-graded sensorial basilar membrane coiled to a snail-shaped form. Its physical basis has recently become well-understood, from a combination of mostly linear hydrodynamics and nonlinear amplifier physics \cite{Dissalbert,KernStoop2003}. To understand how this relates to a discrete network, we focus on the signal pathway. A {\em pure tone} (sound containing only one frequency) arriving at the mammalian cochlea, elicits on the basilar membrane one shallow surface wave that travels down the membrane to its 'resonant place'. Here, the wave becomes strongly  amplified, by frequency-specific nonlinear active units (the so-called outer hair cells). Beyond this point, dissipation by fluidal viscous losses annihilates the wave \cite{Dissalbert,KernStoop2003}. 
The relationship between frequency and place of strongest amplification, measured from the basilar membrane's base, is near-logarithmic. On the logarithmic frequency scale, the interval across which a pure-tone signal is noticeably amplified, is essentially of constant size (Suppl. Mat. 1 a), which suggests a natural partition of the cochlea into sections. These sections provide the {\em nodes} of our network. \begin{figure}[h!!!!!!!!!!!!!!!]
\centering
\includegraphics[width=1\linewidth]{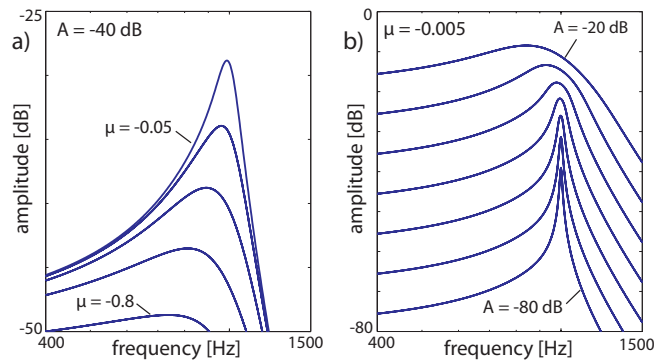}
\caption{Small-signal amplifier characteristics of an isolated node of the cochlear network \cite{KernStoop2003}. Response vs. deviation from the resonant frequency, a) for different distances from bifurcation point ($\mu \in \{-0.05, -0.1, -0.2, -0.4, -0.8 \}$), b) for different signal strengths (steps of 10 dB rms w.r.t. reference level 1). 
}
\label{fig:aplnova_fig5}
\end{figure}

Based on the detailed biophysics and nonlinear dynamics at work in the cochlea \cite{KernStoop2003, Dissalbert}, we developed a model of the sensor consisting of such sections \cite{Martignoli2007,MartignoliStoop2010,Gomez2014}. This model reproduces virtually all mesoscopic biophysical cochlear observations (\cite{Martignoli2007,MartignoliStoop2010,Martignoli2013,Gomez2014,GomezStoop2014}; in particular Suppl. Mat. of Refs. \cite{MartignoliStoop2010,Gomez2014,GomezStoop2014}). Fundamental for this is that the sections share the dynamical properties of the microscopic amplification-providing outer hair cells \cite{Dissalbert,Gomez-Lorimer-Stoop_2015}, which is well-modeled by a stimulated {\em Hopf} process
$$\dot{z}=(\mu+i)\omega_{ch} z - \omega_{ch} |z|^2 z - \omega_{ch} F(t); \: z, \, F(t) \in \mathbb{C},$$
where $z(t)$ denotes the response amplitude, $F(t)$ a stimulation signal, $w_{ch}$ is the characteristic frequency of the Hopf system, and $\mu$ is the {\em Hopf parameter} \cite{Eguiluz2000,Camalet2000,KernStoop2003,Dissalbert,Gomez-Lorimer-Stoop_2015}. At values $\mu<0$, the system is below bifurcation to self-oscillation, but responds towards stimulation signals $F(t)$ as a small-signal amplifier \cite{Wiesenfeld,Stoopbrun}. Dissipation by fluidal viscous losses can be described by tailored 6th-order Butterworth low-pass filters \cite{Martignoli2007,MartignoliStoop2010}. The main characteristics of the isolated node dynamics are collected in Fig. \ref{fig:aplnova_fig5}.
When embedded into a compound cochlea, the response profiles broaden due to the sections' interaction with neighboring ones, reproducing the biological data \cite{Ruggero} extremely well \cite{Martignoli2007}.
The distance of  $\mu$ from bifurcation at $\mu=0$ defines how strongly a node amplifies an incoming signal; we choose this parameter to match the human hearing sensor.
The biophysical properties of the cochlea suggest selecting
the characteristic frequencies of the nodes according to a geometric sequence. We will use a software implementation of an earlier hardware realization of 29 sections or nodes, taking care of 7 octaves ($14.08-0.11$ kHz). Our partition is optimal in the sense that finer partitions yield for the human amplification range, identical results, but coarser partitions lead to distortions in the frequency dependence, if sufficiently strong amplification is required (Suppl. Mat. 2). All nodes will have identical Hopf parameters ({\em 'flat tuning'} $\mu\equiv -0.25$), until we deal with {\em learning}, where we condition the network towards chosen sounds, by tuning unsuited nodes towards weaker amplification \cite{Gomez2014}. 

A node is defined to be {\em 'activated'} by a signal of arbitrary form, if the amplified signal reaches above the psychoacoustic audibility threshold of -50 dB \cite{Smoorenburg1970,GomezStoop2014}, but differing settings of the threshold, e.g. -40 or -60 dB, do not compromise any of the following results. While pure-tone stimulations lead to essentially one activated node, all other (even simple) stimuli lead to complex activation response patterns. In addition to the directly activated nodes (network 'roots'), nodes become activated as the result of the nonlinear interactions among already activated nodes: Interacting 'parent nodes' of frequencies $f_1,\, f_2$ activate 'child nodes' further down the cochlear duct at {\em combination-tone} frequencies $\mid n f_1-m f_2\mid$, $m, n \in \mathbb{N}$. In our network model of hearing, this mechanism contributes the {\em directed links} between activated nodes (Suppl. Mat. 3 and 4). This paradigm contrasts the superposition model of the classical linear frequency analyzer cochlea.

\begin{figure}[!!!!!!!!!!!!!!!!!!!!!!!!!!!h!!!!!!!!!!!!!!!!!!!!!!!!!!!!!!!!!!!!!!!!!!!!!!!!!!!!!!!!!!!!!!!!!!!!!!!!!]
\centering
\includegraphics[width=0.975\linewidth]{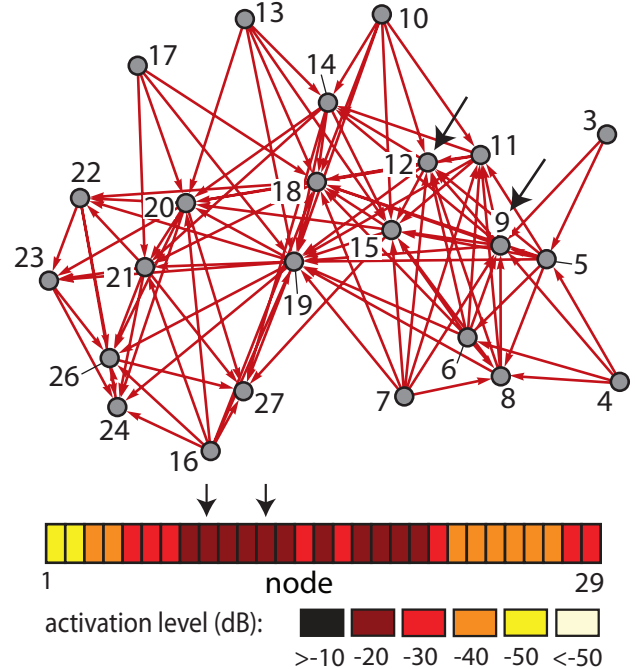}
\caption{Response to stimulation by two complex tones (base frequencies: 2 and 3.35 kHz (black arrows), with 5 harmonics each, both -60 dB strong).
Top panel:  Avalanche activation-network, numbers indicating the location of activated nodes. Origins of incident red arrows indicate 'parents' of activated nodes (cf. video, Suppl. Mat. 4). Lower panel: activity on the unrolled cochlea. For identical input complexity, distinct network sizes are triggered. Cochlea of 29 nodes, covering $(0.11, 14.08)$ kHz logarithmically, $\mu=-0.25$ for all nodes.}
\label{fig1}
\end{figure}

Such {\em stimulation-specific activation-networks} are the focus of this work. Fig. \ref{fig1} exhibits the activation-network from the input of two {\em complex tones} (our complex tones always consist of five consecutive harmonics, starting with the fundamental frequency). Activated nodes relate either to a harmonic of a stimulation, or to a combination-tone of already activated nodes; combination tones have a prominent role in human pitch perception \cite{StoopKern2004, StoopKernPNAS2004,MartignoliStoop2010,GomezStoop2014}. A $0.65$-kHz tone, e.g., is generated as the $2f_1-f_2$ combination tone, with $f_{1,2}=\{2,\,3.35\}$ kHz. Complex activation patterns correspond at the node level to {\em receptive fields} that become increasingly intricate as we move down the cochlea (Suppl. Mat. 1b), reflected in ever more complicated wave forms (Suppl. Mat. 3). For obtaining activation networks, this necessitates inferring from the Fourier spectrum of a node's signal whether a node is a child or a root f (Suppl. Mat. 3). 
While our cochlea is optimized to reproduce all salient features of mammalian hearing, this level of quality is not required. 
The statistical network properties that we focus on are robust regarding sensor variation, as long as combination-tones are sufficiently generated.

We will show that randomly chosen simple inputs generate power-law activation-network size distributions $$ P(s)\sim s^{-\alpha}$$ (where size $s$ is the number of links). This happens autonomously, without the need for additional conditions, such as dynamical synapses or excitatory-inhibitory balance. We will then see how learning drives the network away from power-law behavior. By the close evolutionary and functional relationship between sensory and cortical signal processing \cite{Lorimer2015b}, our results open up a new perspective for understanding generic biological signal processing. On this more general level, our conclusion will be that power-law avalanche-size distributions characterize a fundamental ground-state of biological information processing. 

\begin{figure}[h!!!!!!!!!!!!!!!!!!!]
\centering
\includegraphics[width=0.95\linewidth]{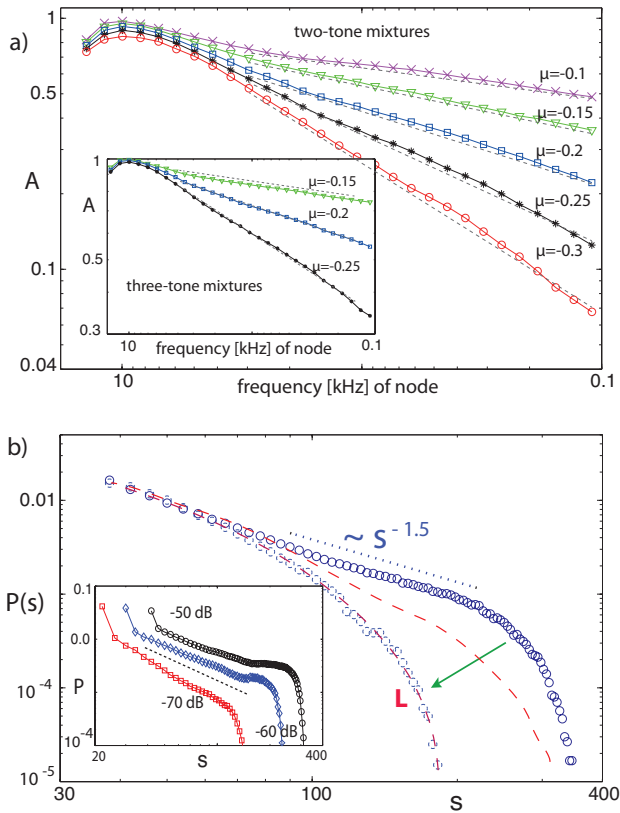}
\caption{Activation power laws. a) Cumulated activity $A$ from 
stimulations by two complex tones of uniformly random input frequencies and random sound levels $\in (-80,-40)$ dB (rms) per tone.
Inset: Results from mixtures of three complex tones. Dotted: maximum likelihood estimates.
b)  Activation-network size $s$ distribution $P(s)$
for mixtures of two complex tones, of randomly chosen intermediate sound levels. Line $L$ and green arrow: Result of learning (see corresponding section; $\mu=-0.25$ (circles), $\mu=-1$ (dashed), $\mu=-2$ (circles and dashed)). Inset: Fixed sound levels (see text).}

\label{fig3} 
\end{figure}

The distribution of the {\em activity} $A$ along the cochlea exhibits the complexity of the interaction among the nodes (computationally, the activation $A$ at node  $j\in\{1,...,29\}$ is defined as $A(j):=\frac{1}{N} \sum_{i=1}^N \Theta(f_i, j)$, where $f_i$ denotes the $i^{\small{\mbox{th}}}$ stimulation, $N$ is the total number of stimulations, and $\Theta(f_i,j)$ is 1 if node $j$ is activated and 0 otherwise). 
Single complex sounds of fundamental frequency uniformly sampled from a non-logarithmic $(0,15)$-kHz frequency interval generate an activities $A$ that still closely follow the sampling power-law ($\alpha =-1$). Mixtures of two and three complex sounds, however, generate $\mu$-dependent power-laws $A \propto f^{-\tilde{\alpha}}$, where $0<\tilde{\alpha}<1$ 
(see Fig. \ref{fig3}a). Because of combination tone dependence on input level, stimulations were randomly chosen from a $(-80,-40)$-dB interval. Regarding the human ear, this corresponds to low to moderate sound levels of 30-70 dB SPL \cite{GomezStoop2014}. Fixing sound level at -60 dB yields identical results.  
The origin of this behavior is clarified by the corresponding activation-network size statistics that share the power-law property (where {\em size} $s$ is defined as the number of activated {\em links}). The exponent and the extension of that power-law regime are governed by two counteracting mechanisms: The level of nonlinearity present ($\mu$) and the input sound level.
Going to higher (lower) sound levels or having $\mu$ closer (further away) from bifurcation both drive the distribution away from the maximal power-law regime. The latter is characterized by exponents 
$\alpha \simeq 1.5$, for the standard $\mu=-0.25$ value and random intermediate sound levels (cf. Fig. \ref{fig3}b). In the figure, dotted lines of slope -1.5 indicate the regions across which a power-law test value $p>0.9$ holds \cite{p-test}. The power-law regime excludes input-specific effects to the left, and finite size effects of the cochlea combined with stimulation specifics to the right. 
These results raise the question whether we deal here with a manifestation of self-organized branching universality with power-law exponent $\alpha=\frac{3}{2}$ \cite{Zapperi,Hergarten}. This particularly since our activation-networks parallel the neuronal avalanches in neocortical networks \cite{Plenz,Plenz2,Beggs3}, and because theoretical models of neocortical networks \cite{Herrmann,Arcangelis,Niebur,Eurich} obtained similar exponents and have emphasized such a connection. On this background, relative to the -50 dB activation threshold, our results at fixed stimulation strengths of  -70, -60, -50 dB could be interpreted as 'underdeveloped', 'fully developed' and 'overdeveloped'. In this order, they would correspond to a transition from subcritical to critical and to supercritical behavior, respectively. The minor bump obtained for -60 dB would then indicate a slightly supercritical behavior or a finite size effect, or a combination of both.

\begin{figure}[h!!!!!!!!!!!!!!!!!!!!!!!!!!!!!!!!!!!!!!!!!!!!!!!!!!!!!!]
\centering
\includegraphics[width=1.02\linewidth]{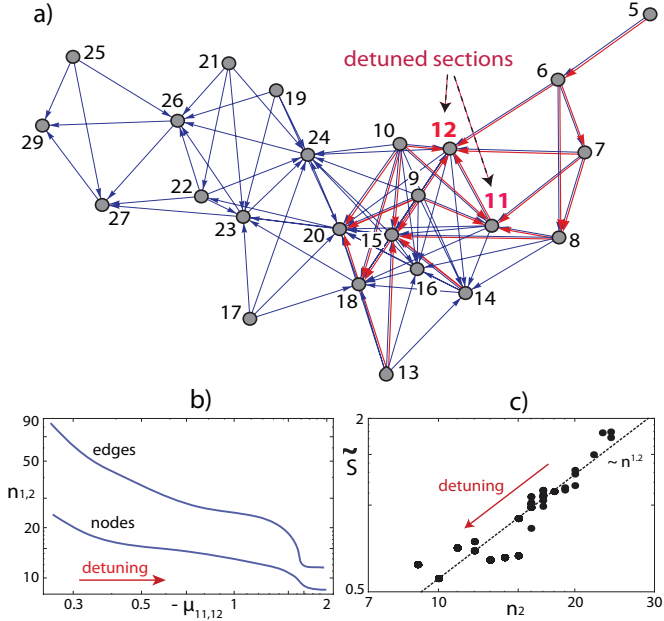}
\caption{Effect of tuning on activation networks. a) Unbiased (blue) vs. tuned (red) cochlea. The red graph is a subgraph of the blue graph. b) Network characteristics edge number $n_1$, node number $n_2$, as a function of the de-tuning of $\mu_{11,12}$. c) Small-worldness $\tilde{S}$ as a function of number of nodes.
}
\label{fig5}
\end{figure}

By scale invariance, power law distributed avalanches express a network's unbiasedness regarding the nature of the information to be processed. To demonstrate that the power-law is lost when we bias the cochlea towards a desired sound, we move the Hopf parameters of sections unrelated to the sound further away from the bifurcation point, a process that extensive usage seems to be made of in mammalian auditory scene analysis \cite{Gomez2014}. Such a {\em 'tuning'} is closely related to learning in neural networks; how learning affects neocortical activity power laws is a topic of strong current interest. 
Already the detuning of two adjacent nodes generates substantial modifications of the elicited activation-networks. If for two complex inputs of fundamental frequencies $f_0^{(1,2)}=\{1.331, 2.120\}$ kHz and five harmonics each (each sound at -70 dB rms), nodes 11 and 12 are jointly gradually detuned from $\mu=-0.25$ to $\mu=-1$, the associated activation-network shrinks substantially (Fig. \ref{fig5}a, and video Suppl. Mat. 5). 
During this process, the associated activation-network size distributions gradually lose their power-law characteristics. This is demonstrated in Fig. \ref{fig3}b for the detuning of two frequency bands (nodes $15,\,16$ and nodes $19,\,20,\,21$) from $\mu=-0.25$ over $\mu=-1.0$ to $\mu=-2.0$. Under this change, the initial power-law distribution $s^{-1.5}$ converts gradually into a strictly convex shape (line $L$), indicating that learning generically drives the network away from power-law criticality. This specific tuning example is generic in the following sense: Substantially simpler tuning patterns do not produce a sound targeting effect sufficient to be qualified as listening, which is co-expressed by the persistence of the power-law nature of $P(s)$. For more information on modalities and real world applications of this learning implementation, readers should consult Ref. \cite{Gomez2014}, from which our example was selected.   

Learning also affects the 'small-worldness' property of our network. To measure 'small-worldness' \cite{smallworldness}, we use the indicators
$\gamma =$ (Clustering coefficient of the considered network) $/$ (Clustering coefficient of the Erd\"{o}s-R\'{e}nyi random network) and $\lambda =$ (av. shortest path-length of the considered network) $/$ (av. shortest path-length of the Erd\"{o}s-R\'{e}nyi  random network). A network is 'small-world' if its 'small-worldness' $\tilde{S}:=\gamma/\lambda$ exceeds unity. For flat tuning, we typically have $\tilde{S}\simeq1.7$ (cf. Fig. \ref{fig5} c). Upon the tuning that changes the activation-network as in Fig. \ref{fig5}a, this value consistently drops, i.e., tuning works against small-worldness. 

\begin{figure}[h!!!!!!!!!!!!!!!!!!!!!!!!!!!!!!!!!!!!!!!!!!!!!!!!!!!!!!]
\centering
\includegraphics[width=0.8\linewidth]{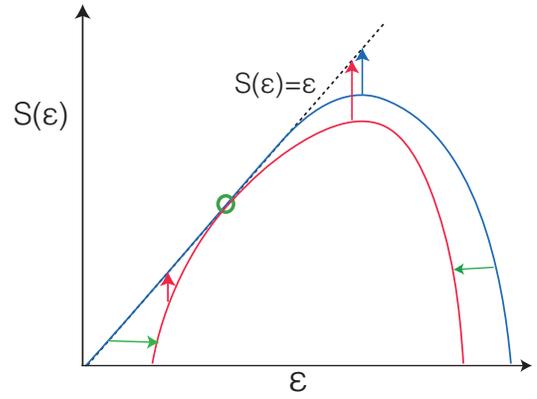}
\caption{Effect of learning on power-law distributions, thermodynamic formalism of  invariant measures (signals) framework \cite{ruelle,stooppeinke,Beck,Gaspardbook}. Blue: Entropy $S(\varepsilon)$ of intermittent systems, power-law characteristics. Red: Entropy of non-power-law behavior, as the result of focusing on a particular measure. The observability $O$ of  an invariant measure $\varepsilon$ decays with time $t$ as $O(\varepsilon,t) \sim e^{-t\,(\varepsilon-S(\varepsilon))}$ (red and blue arrow lengths). Points on the diagonal $\varepsilon=S(\varepsilon)$ refer to measures that decay slower than exponentially. Green arrows and the circle illustrate the learning process. }
\label{fig8}
\end{figure}

There is an information-dynamics argument as to why nature might choose power-law distributions. A system's {\em 'ground-state'} must lack bias towards a particular signal and provide the basis for the system's ability to quickly adapt to new requirements (inputs, tasks) and yet maintain past information for all signals similarly. 
A natural description framework of these properties is within the  '{\em Thermodynamic Formalism}'  \cite{ruelle,stooppeinke,Beck,Gaspardbook}. Focusing on exponential scaling in time $n$, of probability (fractal dimensions) or of support (Lyapunov exponents), e.g., systems are described by a free energy 
$F(\beta)=\lim _{n\rightarrow\infty}\frac{1}{n}\log Z_n(\beta)$ (where $Z_{n}$ is usually a sum of Boltzmann contributions) or by the associated entropy function $S(\varepsilon)=\beta \varepsilon+F(\beta)$, obtained from $F(\beta)$ by a Legendre transform. The probability of observing a value of an observable $\varepsilon$ then scales as \cite{stooppeinke} 
$$P(\varepsilon,n)\,d\varepsilon\sim e^{-n(\varepsilon-S(\varepsilon))}\,d\varepsilon;$$ taking logarithms, we obtain in the thermodynamic limit $\varepsilon=S(\varepsilon)$. While 'normally behaved' systems have 
{\em one} 'observable invariant measure', our power-law distributed invariant density measures admit a continuum of observable invariant measures \cite{stoop complexity} (interpretable as a phase transition). This provides a link to our experimentally observed power-law distributions: Starting from a computational ground-state (power-law characteristics and correlations that decay slower than exponentially), learning forces the system to focus on a particular measure, which destroys the power-law (Fig. \ref{fig8}). Whereas in the ground-state the prediction of the evolution of the system is extremely difficult, for the tuned system this is much simpler  \cite{stoop complexity}, which is co-expressed by an increased computation (measured as the reduction of the complexity of prediction of the system \cite{stoop computation}) after learning. 
Since the specific heat diverges ($\frac{d ^{2}S(\varepsilon)}{d\varepsilon^2}=0)$ and there is no latent heat trace, the ground-state system would indeed be at the critical point. Away from the thermodynamic limit, i.e. for real-world systems, generalized space constraints generically generate power-law deviations that in the thermodynamic limit vanish in a well-controlled manner \cite{lorimerstoop}.

Our approach opens many opportunities to connect with psychophysics and electrophysiology. An example fully in line with our interpretation is anesthesia (known to produce functional disconnection in the posterior complex, causing loss of information capacity by interrupting cortical communication \cite{Alkire_2008}), which pulls the dynamics away from the pre-anesthesia power law \cite{Bellay}.



\begin{thebibliography}{}

\bibitem{Plenz} J.M. Beggs and D. Plenz. 
{\em J. Neurosci.} {\bf 23}, 11167 (2003).

\bibitem{Plenz2} J.M. Beggs and D. Plenz. 
{\em J. Neurosci.} {\bf 24}, 5216 (2004).

\bibitem{Beggs3} T. Petermann, T.C. Thiagarajan, M.A. Lebedev, M.A.L. Nicolelis, D.R. Chialvo, and D. Plenz. 
{\it Proc. Natl. Acad. Sci. U.S.A} {\bf 106}, 15921 (2009).

\bibitem{Boccaletti} S. Boccaletti, V. Latora, Y. Moreno, M. Chavez, and D.-U Hwang. 
{\em Phys. Rep.} {\bf 424}, 175 (2006).
\bibitem{Boccaletti-masterstabilityfunction} M. Chavez, D.-U. Hwang, A. Amann, H.G.E. Hentschel, 
and S. Boccaletti. 
{\em Phys. Rev. Lett.} {\bf 94}, 218701 (2005).

\bibitem{Eurich} C.W. Eurich, J.M. Herrmann, and U. Ernst. 
{\it Phys. Rev. E} {\bf 66}, 066137 (2002).

\bibitem{Herrmann} A. Levina, J.M. Herrmann, and T. Geisel. 
{\em Nat. Phys.} {\bf 3}, 857 (2007). 

\bibitem{Niebur} D. Millman, S. Mihalas, A. Kirkwood, and E. Niebur. 
{\it Nat. Phys.} {\bf 6}, 801 (2010).

\bibitem{Hans} F. Lombardi, H.J. Herrmann, C. Perrone-Capano, D. Plenz, and L. de Arcangelis. 
{\em Phys. Rev. Lett.} {\bf 108}, 228703 (2012).

\bibitem{Dante1}V.M. Eguiluz, D.R. Chialvo, G. Cecchi, M. Baliki, and A.V.  Apkarian. 
{\em Phys. Rev. Lett.} {\em 94}, 018102 (2005).
\bibitem{Haldeman}C. Haldeman and J.M. Beggs.
{\em Phys. Rev. Lett.} {\bf 94}, 058101 (2005).
\bibitem{Arcangelis} L. de Arcangelis and H.J. Herrmann. 
{\it Proc. Natl. Acad. Sci. U.S.A.} {\bf 107}, 3977 (2010).

\bibitem{KernStoop2003} A. Kern and R. Stoop. 
\textit{Phys. Rev. Lett.} \textbf{91}, 128101 (2003).
\bibitem{Dissalbert} A. Kern. \textit{A nonlinear biomorphic Hopf-amplifier model of the cochlea}. PhD Thesis, ETH Z\"urich (2003).
\bibitem{Gomez2014} F. Gomez, V. Saase, N. Buchheim, and R. Stoop. 
\textit{Phys. Rev. Appl.} {\bf 1}, 014003, (2014).
\bibitem{Martignoli2007} S. Martignoli, J.-J. van der Vyver, A. Kern, Y. Uwate, and R. Stoop. 
\textit{Appl. Phys. Lett.} \textbf{91}, 064108 (2007).
\bibitem{MartignoliStoop2010} S. Martignoli and R. Stoop. 
\textit{Phys. Rev. Lett.} \textbf{105}, 048101 (2010).

\bibitem{GomezStoop2014} F. Gomez and R. Stoop. 
{\em Nat. Phys.} {\bf 10}, 530 (2014).


\bibitem{Martignoli2013} S. Martignoli, F. Gomez, and R. Stoop. 
\textit{Sci. Rep.} \textbf{3}, 2676 (2013).

\bibitem{Gomez-Lorimer-Stoop_2015} F. Gomez, T. Lorimer, and R. Stoop. 
{\em Phys. Rev. Lett.}, in press /arXiv (2015).
\bibitem{Eguiluz2000} V.M. Egu\'iluz, M. Ospeck, Y. Choe, A.J. Hudspeth, and M.O. Magnasco. 
\textit{Phys. Rev. Lett.} \textbf{84}, 5232-5235 (2000).
\bibitem{Camalet2000} Camalet, S., Duke, T., J\"ulicher, F. \& Prost, J. 
\textit{Proc. Natl. Acad. Sci. U.S.A.} \textbf{97}, 3183-3188 (2000).
\bibitem{Wiesenfeld} K. Wiesenfeld and B. McNamara. 
{\em Phys. Rev. Lett.} {\bf 55}, 13-16 (1985).
\bibitem{Stoopbrun} B. Derighetti, M. Ravani, R. Stoop, P.F. Meier, E. Brun, and R. Badii. 
{\em Phys. Rev. Lett.} {\bf 55}, 1746-1749 (1985).

\bibitem{Ruggero} M. Ruggero. 
{\em Curr. Opin. Neurobiol.} {\bf 2}, 449 (1992).


\bibitem{Smoorenburg1970} G.F. Smoorenburg. 
\textit{J. Acoust. Soc. Am.} \textbf{48}, 924 (1970).

\bibitem{StoopKern2004} R. Stoop and A. Kern. 
\textit{Phys. Rev. Lett.} \textbf{93}, 268103 (2004).


\bibitem{StoopKernPNAS2004} R. Stoop and A. Kern. 
\textit{Proc. Natl. Acad. Sci. U.S.A.} \textbf{101}, 9179 (2004).

\bibitem{Lorimer2015b}T. Lorimer, F. Gomez, and R. Stoop. 
{\em Sci. Rep.} {\bf 5}, 12492 (2015). 

\bibitem{p-test} A. Deluca and A. Corral. 
{\it Acta Geophys.} {\bf 61}, 1351 (2013).

\bibitem{Zapperi} S. Zapperi,  K.B. Lauritsen, and H.E. Stanley.
\textit{Phys. Rev. Lett.} \textbf{75}, 4071 (1995).

\bibitem{Hergarten} S. Hergarten.
\textit{Phys. Rev. Lett.} \textbf{109}, 148001 (2012).





\bibitem{smallworldness} M.D. Humphries and K. Gurney. 
\textit{PLoS One} {\bf 3}, e0002051 (2008).
\bibitem{ruelle}D. Ruelle. \textit{Thermodynamic Formalism} (Cambridge University Press, 2004).
\bibitem{stooppeinke} J. Peinke, J. Parisi, O.E. R\"ossler, and R. Stoop. \textit{Encounter with Chaos} (Springer, 1992).
\bibitem{Beck} C. Beck, and F. Schl\"ogl. \textit{Thermodynamics of Chaotic Systems: An Introduction} (Cambridge University Press, 1995).
\bibitem{Gaspardbook} P. Gaspard. \textit{Chaos, Scattering and Statistical Mechanics} (Cambridge University Press, 2005). 
\bibitem{stoop complexity} R. Stoop, N. Stoop, and L. Bunimovich. 
{\em J. Stat. Phys.} {\bf 114}, 1127 (2004).
\bibitem{stoop computation}R. Stoop and N. Stoop. 
{\it Chaos} {\bf 14}, 675 (2004).

\bibitem{lorimerstoop}T. Lorimer, F. Gomez, and R. Stoop. 
{\it Sci. Rep.} {\bf 5}, 12492 (2015).


\bibitem{Alkire_2008} M.T. Alkire, A.G. Hudetz, and G. Tononi. 
{\it Science} {\bf 322}, 876 (2008). 

\bibitem{Bellay} T. Bellay, A. Klaus, S. Seshadri, and D. Plenz. 
{\it eLife} {\bf 4}, e07224. DOI: 10.7554 (2015).

\end{thebibliography}
\end{document}